\documentclass[preprint]{aastex}

\shortauthors{Batalha et al.}

\begin{document}


\title{Information Content Analysis for Selection of Optimal \\ JWST Observing Modes for Transiting Exoplanet Atmospheres}

\author{Natasha E. Batalha\altaffilmark{1,2}}
\affil{Department of Astronomy \& Astrophysics, Pennsylvania State University, State College, PA 16802}
\email{neb149@psu.edu}
\and

\author{M.R. Line}
\affil{School of Earth \& Space Exploration, Arizona State University, Phoenix, AZ 85282}

\altaffiltext{1}{Center for Exoplanets and Habitable Worlds, Pennsylvania State University, State College, PA 16802}
\altaffiltext{2}{Planetary Systems Laboratory, Goddard Space Flight Center, Greenbelt, MD 20770}

\begin{abstract}
The James Webb Space Telescope (JWST) is nearing its launch date of 2018, and is expected to revolutionize our knowledge of exoplanet atmospheres. In order to specifically identify which observing modes will be most useful for characterizing a diverse range of exoplanetary atmospheres, we use an information content based approach commonly used in the studies of Solar System atmospheres. We develop a system based upon these information content methods to trace the instrumental and atmospheric model phase space in order to identify which observing modes are best suited for particular classes of planets, focusing on transmission spectra. Specifically, the atmospheric parameter space we cover is T=600-1800 K, C/O=0.55-1, [M/H]=1-100$\times$Solar for a R=1.39 R$_J$, M=0.59 M$_J$ planet orbiting WASP-62-like star. We also explore the influence of a simplified opaque gray cloud on the information content. We find that obtaining broader wavelength coverage over multiple modes is preferred over higher precision in a single mode given the same amount of observing time. Regardless of the planet temperature and composition, the best modes for constraining terminator temperatures, C/O ratios, and metallicity are NIRISS SOSS+NIRSpec G395. If the target's host star is dim enough such that the NIRSpec prism can be used, then it can be used instead of NIRISS SOSS+NIRSpec G395. Lastly, observations that use more than two modes, should be carefully analyzed because sometimes the addition of a third mode results in no gain of information. In these cases, higher precision in the original two modes is favorable. 

\end{abstract}


\section{Introduction}
The James Webb Space Telescope (JWST) is equipped with eleven different observation modes across eight different wavelength ranges and six different spectral resolving powers that can all be used for transmission spectroscopy of exoplanets. While several studies have sought to identify what the limits of these modes will be, in terms of exoplanet characterization \citep{bei14,bar15a,bar15b,bat15,gre16,roc16}, little work has been done to specifically identify which instrument modes or combinations of modes will be most useful for characterizing a diverse range of exoplanetary atmospheres. The most rigorous way of accomplishing this is through atmospheric retrieval, which links atmospheric models to the data in a Bayesian framework \citep{lin12,lin13a,lin14,lin15,mad12,ben12,ben13,wal15}.

For example, \citet{bar15a} simulated an observation with NIRSpec prism and MIRI Low Resolution Spectrometer (LRS) for four specific case studies: a hot Jupiter, a hot Neptune, GJ 1214, and Earth. For each observation they performed a full retrieval analysis to determine the prospects for identifying the true atmospheric state of the planet and assessed the possible effects of star spots and stitching on the results. \citet{gre16} also simulated the observations of four specific planet archetypes (hot Jupiter, warm Neptune, warm sub-Neptune, cool Super-Earth) in three different combinations of modes: NIRISS Single Object Slitless Spectrograph (SOSS) only, NIRISS+NIRCam, NIRISS+NIRCam+MIRI. They concluded that spectra spanning 1-2.5$\micron$ will often provide good constraints on the atmospheric state but that in the case of cloudy or high mean molecular weight atmospheres a 1-11$\micron$ spectrum will be necessary. While both of these studies offer insights into the kind of data we will obtain from JWST, they do not yet simulate observations for a diverse instrument phase space for a wide variety of planet types, mainly driven by the computational limits of Markov chain Monte Carlo (or related) methods. Evaluating a wider range of planet types and instrument modes, however, is necessary if we want to be able to optimize our science output with JWST. 

To solve this problem, we use information content (IC) analysis, commonly used in studies of Earth and Solar System atmospheres. As some Earth examples, \citet{kua10} used IC analysis to determine the 20 best channels from each CO$_2$ spectral region for retrieving the most precise CO$_2$ abundance measurements obtained with the Orbiting Carbon Observatory. Using the channels selected from their analysis, they were able to achieve precision better than 0.1 ppm. Similarly, \citet{sai09} demonstrated that separately selecting a subset of the 15-$\micron$ CO$_2$ channels based on IC analysis, yielded the same precision on their retrieved results as the entire 15-$\micron$ band. 

IC analysis has also been used with exoplanet spectra. \citet{lin12} quantified the increase in information content that comes from an increase in signal to noise and spectral resolving power for an arbitrary wavelength range, 1-3 $\micron$, and for a single planet case. 

Most recently\footnote{We became aware of their publication only a few days prior to our submission.}, \citet{how16} also presented IC analysis as a way to optimize JWST observations with a simple three-parameter model (uniform temperature, uniform metallicity, and opaque cloud deck). They designed 7 different theoretical JWST programs, which range in length from observing a single mode in a single transit to observing nine modes across nine transits. They test these programs on a target list composed of 11 targets all with T$_{eq}>958$~K, which were separated into bright (J$<$8), medium (8$<$J$<$11), and dim (J$>$11) targets. They find that MIRI LRS in slitless mode provides more information than in slit mode. They also found that NIRISS SOSS consistently provides the most information content for a given integration time compared with other observing modes. Finally, they conclude that IC analysis is a powerful technique that can be used for selecting JWST observing modes. 

Here we expand this analysis to look at a broader range of planet archetypes ranging from T$_{eq}$=600-1800 K, C/O=0.55-1, [M/H]=1-100$\times$Solar, with no cloud/grey cloudy/haze (84 planet types), across every possible combination of JWST transit spectroscopy modes. In doing so we answer the following questions: 
\begin{enumerate}
    \item Is there a mode or combination of modes that will provide more information per unit observing time?
    \item And, does that differ across different combinations of C/O ratio, [M/H], or temperature?
    \item Is it better to sacrifice wavelength coverage across several different modes or to increase the precision of a single mode? 
    \item Is there a point where the addition of more observing modes stops tightening the constraints on the retrieved model parameters? 
\end{enumerate}

In \S2 we explain the theory of IC analysis, in \S3 we explain the transmission forward models and our JWST noise estimates. In \S4, we look at the results of the IC analysis and verify the results using a retrieval algorithm. In \S5 we discuss these results and end with concluding remarks in \S6. 

\section{Information Content} 
In atmospheric retrieval \citep{cha68,rod76,rod00,two77} the goal is to obtain the most likely set of model parameters given a set of observations. The model parameters, which define the atmospheric state, are just a vector, \textbf{x}, of length \emph{n}, that is usually composed of gas mixing ratios, temperature at each atmospheric level, and any other atmospheric parameters pertinent to the model. The relationship between \textbf{x} and the observations is given by
\begin{equation}
    \mathbf{y = F(x_a) + K(x-x_a)}
\end{equation}
where \textbf{F(x)} is the model and \textbf{x}$_\mathbf{a}$ is the initial guess of the true state (also known as the prior). \textbf{K} is an $m \times n$ Jacobian matrix given by
\begin{equation}
    K_{ij} = \frac{\partial F_i(\mathbf{x})}{\partial x_j}
\end{equation}
The Jacobian describes how sensitive the model is to perturbations in each state vector parameter at each wavelength position. $F_i$ is the measurement in the $i^{th}$ channel and $x_j$ is the $j^{th}$ state vector parameter.  The jacobians are numerically computed with a centered-finite difference scheme. 

We assume that an exoplanet transit transmission spectrum can be fully described with a 4 parameter state vector: $\mathbf{x}=[T, C/O, [M/H], \times R_p]$, where $T$ is the isothermal terminator temperature, $C/O$ is the carbon to oxygen ratio, $[M/H]$ is the log-metallicity relative to solar, and $\times R_p$ is a scaling factor to the reported radius arbitrarily defined at 10 bars. This is almost certainly an overly simplistic description as there are numerous additional processes at play such as atmospheric dynamics, photochemistry, clouds etc. Certainly this analysis can be extended for arbitrary atmospheric descriptions.  

The information content, measured in bits, quantitatively describes how the state of knowledge (relative to the prior) has increased by making a measurement \citep{sha62,lin12}. It is computed as the reduction in entropy of the probability that an atmospheric state exists given a set of measurements: 
\begin{equation}
    H = entropy(P(\mathbf{x}))-entropy(P(\mathbf{x|y}))
\end{equation}
where 
\begin{equation}
    P(\mathbf{x}) \propto e^{-0.5\mathbf{(x-x_a)}^T\mathbf{S_a^{-1}(x-x_a)}}
\end{equation}
\begin{equation}
    P(\mathbf{x|y}) \propto e^{-0.5 J(\mathbf{x})}
\end{equation}
In (4) $\mathbf{S_a}$ is a $n \times n$ \emph{a priori} covariance matrix, which defines the prior state of knowledge, e.g., the uncertainties on the atmospheric state vector parameters before we make a measurement. $J(\mathbf{x})$ is the cost function which is given by: 
\begin{equation}
    J(\mathbf{x}) = \mathbf{(y-Kx)}^T\mathbf{S_e}^{-1}\mathbf{(y-Kx)} + \mathbf{(x-x_a)}^T\mathbf{S_a^{-1}(x-x_a)}
\end{equation}
where $\mathbf{S_e}$ is the $m \times m$ data error covariance matrix. The first term in the cost function describes the data's contribution to the state of knowledge (``chi-squared") and the second describes the contribution from the prior. 

Assuming Gaussian probability distributions \citep{rod00} the information content can be written in terms of the posterior and prior covariance matrices: 
\begin{equation}
    H = \frac{1}{2} \ln (|\mathbf{\hat{S}}^{-1}\mathbf{S_a}|)
\end{equation}
and 
\begin{equation}
    \mathbf{\hat{S}} = (\mathbf{K^TS_e^{-1}K + S_a^{-1}})^{-1}
\end{equation}
where $\mathbf{\hat{S}}$ is the posterior covariance matrix that describes the uncertainties and correlations of the atmospheric state vector parameters after a measurement is made.

As an illustrative example relevant to  JWST, if we were only interested in deciding between NIRISS, NIRCam, and MIRI to maximize the total retrievable information (let's say: T, $\times$R$_p$, C/O, [M/H]), the goal would be to minimize the elements of $\mathbf{\hat{S}}$. Because there will likely be little prior knowledge on these parameters, $\mathbf{S_a}^{-1}<<\mathbf{K^TS_e^{-1}K}$, the mode covering wavelengths with the greatest sensitivity to each of the state vector parameters (maximum values of $\mathbf{K}$) and the smallest error (minimum values of $\mathbf{S_e}$), will have the lowest values of $\mathbf{\hat{S}}$. The relative information content from one mode to the next under these assumptions will be largely independent of $\mathbf{S_a}$ in Eqn. 7.  The mode with the highest value for \emph{H} will yield the most information of the atmospheric state, and would thus be considered the optimal mode.
%%%%%%%%%%%%%%%%%%%%%%%%%%%%%%%%%%%%%%%%%%%%%%%%%%%%%%%%%%%%%%%%%%
\section{Modeling \& Retrieval Approach}
\subsection{Transit Transmission Spectra Models \& their Jacobians}

We use the {\it chemically-consistent} transit transmission approach described in \citet{kre15}. Given the temperature-pressure profile of the atmosphere and the elemental abundances parametrized with metallicity, [M/H], and C/O, the model first computes the thermochemical equilibrium molecular mixing ratios (and mean molecular weight) using the publicly available Chemical Equilibrium with Applications code (CEA, \citet{mcb96})\footnote{https://www.grc.nasa.gov/WWW/CEAWeb/}. The thermochemically derived opacity relevant mixing ratio profiles (H$_2$O, CH$_4$, CO, CO$_2$, NH$_3$, H$_2$S, C$_2$H$_2$, HCN, TiO, VO, Na, K, FeH, H$_2$, He), temperature profile, cloud and haze proprieties, and planet bulk parameters (10 bar radius, stellar radius, planetary gravity) are then fed into a transit transmission spectrum model (\citet{lin13b,gre16,lin16}, using the \citet{fre08,fre14} opacity database) to compute the wavelength-dependent eclipse depth at the appropriate instrument spectral resolving power. For cloudy simulations, we assume a hard gray cloud top pressure set to be at the 1 mbar pressure level, below which the transmittance is set to zero and use the ``Rayleigh Haze" power law parameterization \citep{lec08} to describe hazes.

This simplistic treatment of clouds and hazes is motivated by WFC3+STIS observations \citep{kre14b, knu14,sin16}, in which simple gray cloud top pressures, and power law parameterizations are sufficient to fit the data. Additionally, more complex cloud model parameterizations are not suitably motivated by the data and our generally poor understanding of the very complex coupled, 3D-dynamical-radiative-microphysics in non-Earth-like planets \citep[e.g.][]{lee15}. A 1 mbar pressure level was chosen arbitrarily so that the absorption features as viewed in transmission were muted, but not completely masked, as demonstrated in \citep{iye16}. While perhaps an overly simplistic cloud model, it still allows us to assess which modes are the most susceptible to a loss of information content because of the presence of clouds. We emphasize that the goal is not to identify the ideal setup for characterizing clouds and hazes, but rather the influence that clouds of some form can have on our ability to extract other useful quantities. We discuss this further in Section 5.3.

For all initial state vectors, \textbf{x}, we assume a planet radius of R=1.39 R$_J$ and mass of M=0.59 M$_J$ around WASP-62 (T$_{eq}$=6230.0~K, F7, 1.28~R$_\sun$). The WASP-62 system was chosen because it was identified as a potential target for the JWST Early Release Science \citep{ste16} and because it has a magnitude that does not saturate the instrument modes explored here (J=9.07). \citet{how16}, in contrast, did explore ranges in stellar magnitudes and found that the NIRSpec prism, which not explored here, is the best mode for faint sources.

We do not explore parameter space in planet radius and mass, because changes in radius  will affect the spectrum uniformly in wavelength space and we assume that the mass will be known for planets we observe with JWST. Changing the star will affect the error profiles because of the different SED peaks and the different stellar magnitudes. These effects will be minor compared to the effects that come from changing the planetary atmosphere parameters. Therefore, we fix the stellar type as well. 

We explore 7 temperatures ranging from T$_{eq}$ = 600-2000 K, 2 C/O ratios (0.55 and 1) and two metallicities (1 and 100$\times$Solar). The ranges in C/O and [M/H] were chosen to represent a diverse set of chemical compositions \citep{mad12,kre14}. In contrast, \citet{how16} is limited to T$_{eq}>$958~K and do not explore different C/O ratios. We explore three different cloud scenarios: no clouds, grey cloud, and power-law haze. For each of these 84 combinations of planet types, we compute a \emph{separate} Jacobian. 

\begin{figure}
\centering
 \includegraphics[angle=270,width=6in]{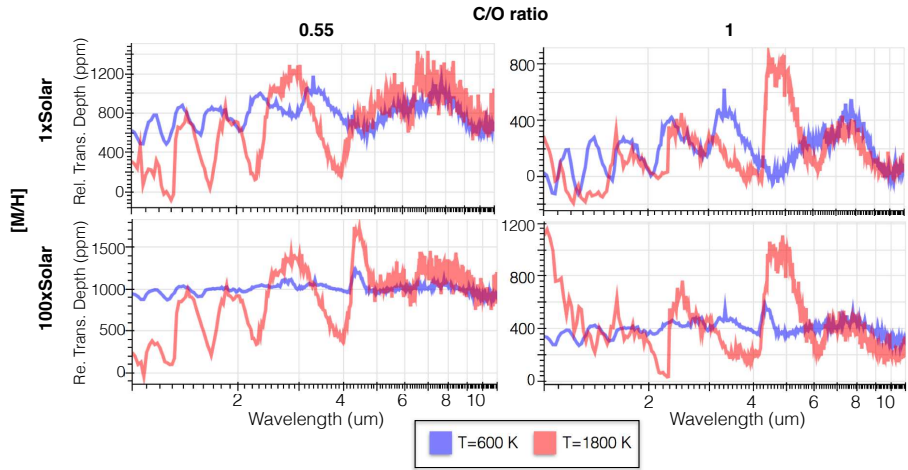}
\caption{Eight transmission spectra forward models used for the IC analysis. Each transmission spectrum is computed for a WASP-62 type star with a planet the size and mass of WASP-62 b. Atmospheres were computed assuming chemical equilibrium with the specified temperature (red: 1800 K, blue: 600 K). Each subpanel represents a different combination of C/O and [M/H], as labeled. \label{fig1}}
\end{figure}

\begin{figure}
\centering
 \includegraphics[angle=90,width=6in]{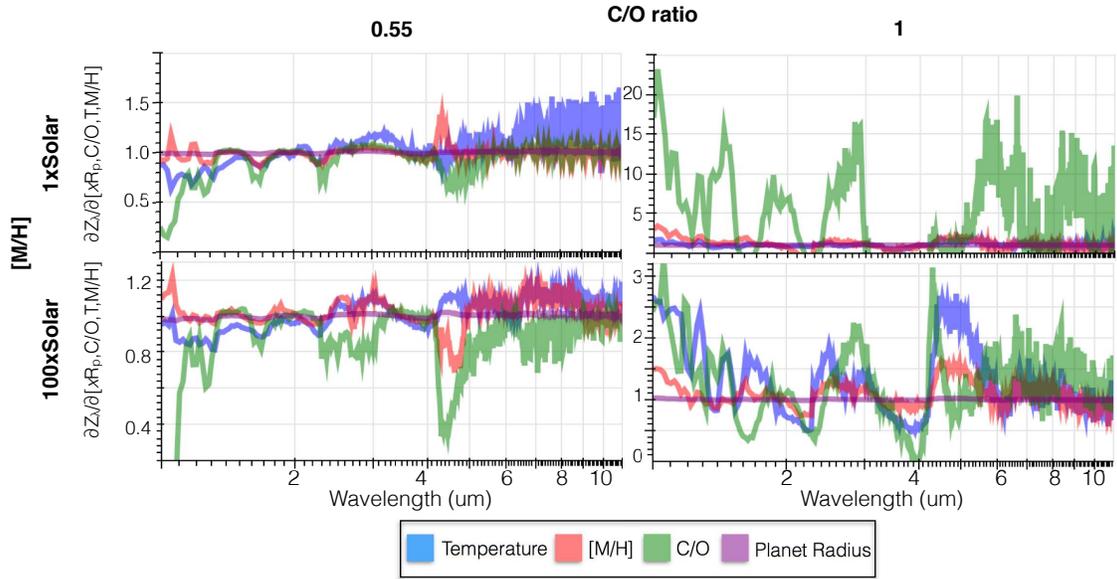}
\caption{The elements of the Jacobian (Eqn. 2) for the T=1800 K transmission spectrum forward models  cases shown in Fig \ref{fig1}. Each Jacobian is binned to R=100. \label{fig2}}
\end{figure}

We choose eight representing planet types to display our results: T = 1800 K (Figure \ref{fig1}, red) and T = 600 K (Figure \ref{fig1}, blue) with C/O=0.55 and 1, and with [M/H]=1$\times$Solar and 100$\times$Solar. Figure~\ref{fig2} shows the Jacobian for different C/Os and [M/H]'s at T=1800 K (at a resolving power, R=100). Because each combination of C/O and [M/H] have very different Jacobians, instrument mode selection must be optimized without making assumptions about the atmospheric composition of the planet \emph{a priori}. 

%%%%%%%%%%%%%%%%%%%%%%%%%%%%%%%%%%%%%%%%%%%%%%%%%%%%%%%%%%%%%%%%%% 
\subsection{JWST Noise Models}
To create noise simulations we use the beta version of Space Telescope Science Institute’s online\footnote{https://devjwstetc.stsci.edu} Exposure Time Calculator. Therefore, noise estimates are expected to change as the final release date approaches as it \emph{has not yet been verified against instrument team's calculations}. The JWST ETC includes the most up-to-date estimates for background noise, PSFs, wavelength-dependent instrument throughputs, optical paths, and saturation levels. It \emph{does not} include estimates of residual systematic noise floors. It currently supports all officially-supported JWST observing modes and provides one-dimensional calculations of flux and background rates (e$^-$/s) per resolution element, $F_{\lambda}$ and $B_{\lambda}$, respectively.

To calculate the duty cycle, $d$, we must determine the total number of allowable groups per integration, $n_{grp}$, before detector saturation. In JWST's MULTIACCUM readout scheme, a group is a set of consecutively read frames separated by a reset frame. All exoplanet observations will have a single frame per group. Therefore, to maximize the duty cycle we maximize the number of groups per integration. The number of groups is related to the duty cycle via the relation, $d = \frac{n_{grp}-1}{n_{grp}+1}$. We step up the number of groups in the JWST ETC until we reach a point of saturation for each instrument mode. For WASP-62-like target, which has a J-mag of 9.07, we calculate duty cycles that are all $>$75\% for all the instrument modes. We do not compute noise simulations for the NIRSpec prism because of saturation limitations. 

With the number of groups, the total shot noise on the star can be computed via:
\begin{equation}
\sigma_{shot,\lambda}^2 = F_{in,\lambda}*t_{in}+F_{out,\lambda}*t_{out}
\end{equation}
Where the $F_{out,\lambda}$ is the count rates computed from the JWST ETC for WASP-62 in e$^{-}$/s, and  $F_{in}=F_{out,\lambda}(1-R_{p,\lambda}^2/R_{*,\lambda}^2)$. $R_{p,\lambda}^2/R_{*,\lambda}^2$ is calculated from the model outlined in $\S$3.1. $t_{in}$ is total transit duration (2 hours) multiplied by the duty cycle, which gives the total on source-time. We assume that equal time is spent observing in and out of transit so that $t_{in}=t_{out}$. Therefore, we assume the light curve is equivalent to a step function and do not fully model the ingress and egress.

We also include readnoise and background noise so that the total noise is:
\begin{equation}
    \sigma_{tot,\lambda}^2 = \frac{1}{(F_{out,\lambda}t_{out})^2}(\sigma_{shot,\lambda}^2 + B_\lambda (t_{out}+t_{in}) + RN^2 n_{pix} n_{int})
\end{equation}
Where $B_\lambda$ is the count rate of the background (observatory background+dark current) in e$^{-}$/s, computed with the JWST ETC, $n_{pix}$ is the number of extracted pixels and $n_{int}$ is the number of integrations in the entire transit observation (4 hours). $RN$ is the readnoise. We use $RN$=18e$^-$ for the HgCdTe detectors (NIRISS, NIRSpec, and NIRCam), and $RN=28$e$^-$ for MIRI \citep{gre16}. Eqn. 10 is used to create the error covariance matrices, $\mathbf{S_e}$ in Eqn. 8.
%%%%%%%%%%%%%%%%%%%%%%%%%%%%%%%%%%%%%%%%%%%%%%%%%%%%%%%%%%%%

%%%%%%%%%%%%%%%%%%%%%%%%%%%%%%%%%%%%%%%%%%%%%%%%%%%%%%%%%%%%
\section{IC Analysis Results}
\begin{figure}
\centering
 \includegraphics[angle=270,width=6in]{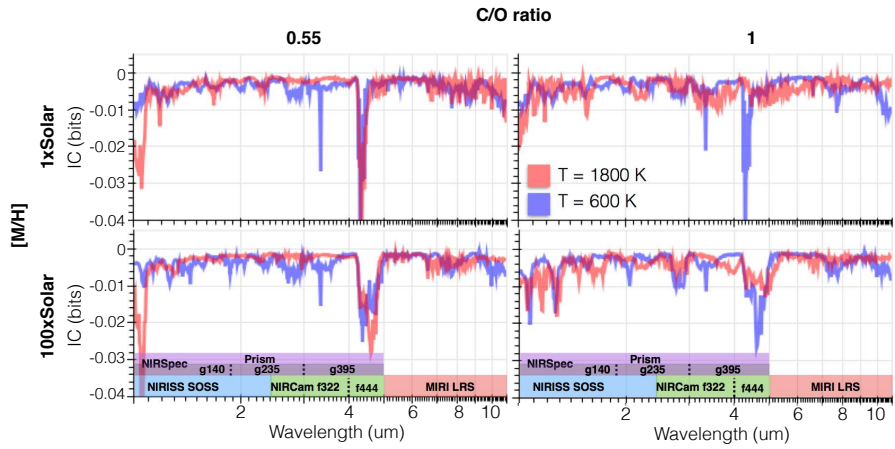}
\caption{Loss of information content as a result of the removal of a R=100 element of an observation of a spectrum with a uniform precision of 1 ppm. Each panel contains a different [M/H] and C/O, equivalent to Fig \ref{fig1}. Also as with Figure~\ref{fig1}, red lines are for a simulation of a target with T=1800 K and blue lines are for a simulation of a target with T=600 K. Shaded boxes in bottom panels shows the available transit timing spectroscopy modes on board JWST. \label{fig3}}
\end{figure}

Before we compute the IC of certain instrument modes, we can predict what wavelength regions are going to hold the most information. This will give us intuition for why certain modes are better suited for constraining atmospheric parameters than others. To do this, we start by computing the IC of a synthetic observation of R=100, assuming a precision of 1 ppm, across the full JWST wavelength region 1-12$\micron$. Then, we sequentially remove each R=100 bin from the spectrum and recompute the IC. Figure \ref{fig3} shows the loss of IC from the removal of each R=100 element for representative temperature-C/O-metallicity combinations. Regardless of temperature and chemistry, the removal of spectral elements near 1-1.5$\micron$ and 4-4.5$\micron$ always results in the greatest loss of IC.  

The only mode that contains both of these wavelength regions is the NIRSpec prism. The prism will undoubtedly be widely used for exoplanet spectroscopy however, it has a high saturation limit (J$<$10.5) and a low spectral resolving power of R$\sim$100. We constrain our future discussion to the modes shown in Table \ref{tab1} because we assume that the observer will usually pick the prism for transmission spectroscopy if their target is dim enough and if a low spectral resolving power is sufficient for retrieving their desired atmospheric parameters. 

Modes that only contain 1-1.5$\micron$ include NIRSpec G140M/H, and NIRISS SOSS. Modes that only contain 4-4.5$\micron$ are NIRCam F444, and NIRSpec G395M/H. While this is informative, it does not directly dictate which modes hold the highest IC. Next, we compute, for each individual mode, the increase in information content as a function of increasing signal-to-noise.

\subsection{Single Mode Analysis} 
Computing the increase in IC as a function of precision on the planet spectrum will tell us how much information we are losing for each instrument mode as the precision on the spectrum decreases. This decrease is usually the result of less observing time, a dimmer target, stellar variability or instrument systematics/noise floors. For each calculation we assumed there were constant errors across the wavelength range of the instrument mode. This assumption is \emph{only} used in the section in order to explore the exact relationship between the error on the retrieved parameters and the error on a spectrum. In reality, all JWST instruments will have highly wavelength-dependent throughputs with sharply declining edges. To minimize the effects of this on this particular analysis we do not use the edges of the observation bands where the noise increases sharply. Lastly, the conclusions made in this section are later verified by full noise simulations.

\begin{figure}
\centering
 \includegraphics[angle=270,width=6in]{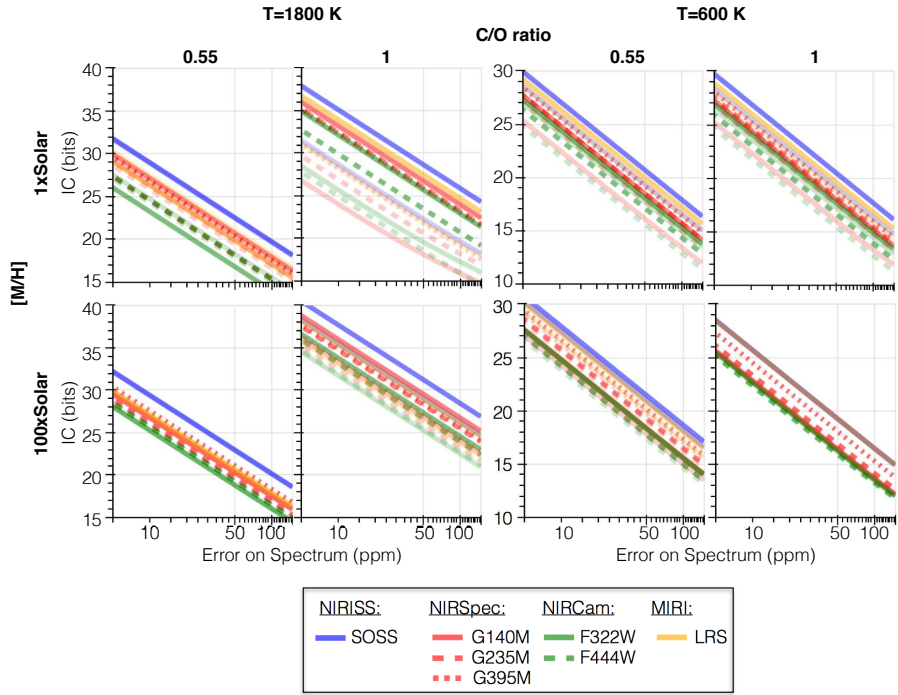}
\caption{The information content as the precision on a spectrum increases. The colored lines in each panel represents a different observing mode. Each panel represents a different combination of [M/H] and C/O for a T=1800 K (left) or T=600 K (right) target. Noise simulations were not computed for these calculations. Instead, the "error on spectrum" dictates a fixed precision over the wavelength region of the particular observing mode. Transparent lines show, for the same observation mode, the loss in information content with the addition of a grey cloud at 1 millibar.\label{fig4}}
\end{figure}

Each panel in Figure~\ref{fig4} shows this function for a different chemical configuration and temperature. The opaque lines are calculations with no clouds and the transparent lines are calculations with a grey cloud at 1 mbar. Interpreting these plots is done best by picking a fixed IC and looking to see what precision each mode needs in order to attain that level of IC. For example, for 1$\times$Solar, C/O=0.55, the observer needs a high precision of 7 ppm with NIRCam F322 to achieve an IC=25 bits. If the observer were to choose NIRISS SOSS instead, to achieve IC=25 bits, they would only need to achieve a precision of 30 ppm. 

In all four temperature, chemical configurations, and cloud assumptions, NIRISS SOSS (purple) has the highest IC and NIRCam F322W2, and F444W (green) have the lowest. The simplest way to understand this is by analyzing Figure~\ref{fig2}. NIRISS SOSS covers a large wavelength space specifically over points that hold a lot of information (strong absorption features). The latter statement is especially important. MIRI LRS, for example, covers an even larger wavelength space but over points that hold less information than those in NIRISS SOSS (weaker absorption features). Because of MIRI LRS' larger wavelength coverage, it holds the second highest IC content in most temperature/chemistry configurations. We discuss these results further in the \S5. 

Adding in clouds (transparent lines) decreases the total IC of each mode and in some cases, changes their relative ranking. In general, shorter wavelength modes lose more IC because of clouds as opposed to than longer wavelength modes. This is clearly seen in the case of Figure~\ref{fig4} T=1800 K, C/O=1 \& 1$\times$Solar. NIRSpec G140M/H loses up to $\sim$10 bits of information with the addition of clouds whereas the other modes only lose $\sim$5 bits. This is as expected, a grey opacity source will strongly mute absorption features at shorter wavelengths ($<2 \micron$), while longer wavelengths will be less affected. NIRSpec G140M/H covers 1-2 $\micron$. NIRISS SOSS is less susceptible to this loss of IC because of its extended coverage out to 2.8 $\micron$. 

Although it is useful to compute the IC of each mode for generic mode comparison purposes, it does not directly tell us what the precisions on the desired parameters (T, C/O, [M/H]) will be. For this we can look at the diagonal elements of the matrix $\mathbf{\hat{S}}$, Eqn. 8, which we show below are an accurate enough description of the parameter uncertainties . 

\subsection{Validation of Covariance Matrix Approximation Against Full Retrievals}
\begin{figure}
\centering
 \includegraphics[angle=270,width=6in]{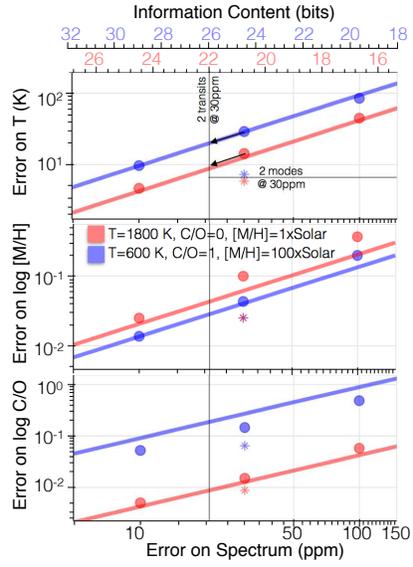}
\caption{The red and blue curves show the derived errors from the IC analysis on the state parameters of interest for an observation with only NIRISS of a planet with 1) T=1800 K, C/O=0.55 and [M/H]=1$\times$Solar, and no cloud and 2) T=600 K, C/O=1 and [M/H]=100$\times$Solar, and no cloud. The errors on each parameter were calculated using Eqn. 8. The top x axis shows IC content (taken from y axis in Figure~\ref{fig4}). Circles show the errors on the parameters as a result of a multinest retrieval scheme. Asterisks show the errors on the parameters as a result of using a multinest retrieval scheme on a combined NIRISS \& NIRSpec G395M observation.\label{fig6}}
\end{figure}

A potential issue in our analysis is the assumption of the Gaussianaity of the posterior, described via the covariance matrix $\mathbf{\hat{S}}$ (e.g., see \citet{ben12} for a discussion). This can, for instance, happen when a parameter is not constrained other than an upper limit. The covariance matrix approximation within the optimal estimation framework would try to approximate such upper limits with a Gaussian, possibly over, or underestimating the uncertainties.  However, here we validate this assumption in our atmospheric parametrization and data regimes by performing full retrievals (e.g., \citet{ben13,lin13a,wal15,lin16}) using \emph{PyMultiNest} \citep{buc16} on select cases.  These results are summarized in Figure~\ref{fig6}. The circles in Figure~\ref{fig6} show the errors on the parameters as a result of the \emph{PyMultiNest} retrieval described above. They indicate that the covariance matrix estimated error analysis and the full retrieval generally agree for the relevant parameters over a wide range of spectral uncertainties. 

As an illustrative example, the curves in Figure~\ref{fig6} show the relationship between parameter precision, data error, and information content for NIRISS SOSS for the case of T=1800 K, C/O=0.55 \& 1$\times$Solar (red) and T=600 K, C/O=1 \& 100$\times$Solar (blue). For both cases, a loss in 10 bits of information translates to a factor of $\sim$10 increase in the error on the atmospheric parameters (i.e. for T=1800 K, $\sigma_T=4$ with 30 bits of IC and $\sigma_T=40$ with 20 bits of of IC). This result can be readily understood by inspecting Eqn. 7. If uncertainties on all parameters decrease by 10, then the diagonal elements of $\mathbf{\hat{S}}$ decrease by a factor of 100. The determinant of the 4$\times$4 $\mathbf{\hat{S}}$ matrix then increases by a factor of $10^8$, where one half of the natural log of $10^8$ is 9.  As another reference point, an IC difference of $\sim$3 corresponds to a factor $\sim$ 2 improvement in the constraints on each of the 4 parameters.

Also shown in Figure~\ref{fig6} (with asterisks) are \emph{PyMultiNest} retrievals for the case of an observation with NIRISS SOSS \emph{and} NIRSpec G395M. This reveals that expanding wavelength coverage to two observation modes results in a lower error than two observations in the same mode. The error on the temperature of a 30ppm observation with NIRISS and NIRSpec G395M (horizontal line) is less than the error on the temperature of a $30/\sqrt(2)$ppm (vertical line) observation with only NIRISS SOSS. Additionally, a factor of $\sqrt(2)$ improvement in the noise is only valid in the case of little systematic noise. In the \S5.2 we explore this further. 

%%%%%%%%%%%%%%%%%%%%%%%%%%%%%%%%%%%%%%%%%%%%%%%%%%%%%%%%%%%
\subsection{Two Mode Analysis}
Not counting modes with overlapping wavelength space, there are approximately 8!$\sim$40,320 combinations of modes on board JWST for exoplanet spectroscopy alone. In order to narrow down these combinations, we first focus on only two mode comparisons in transmission, for the two temperature cases. Instead of fixing the noise as was done in $\S$4.1, we compute full noise models for a single transit in each mode. 

\begin{figure}
\centering
 \includegraphics[angle=270,width=5in]{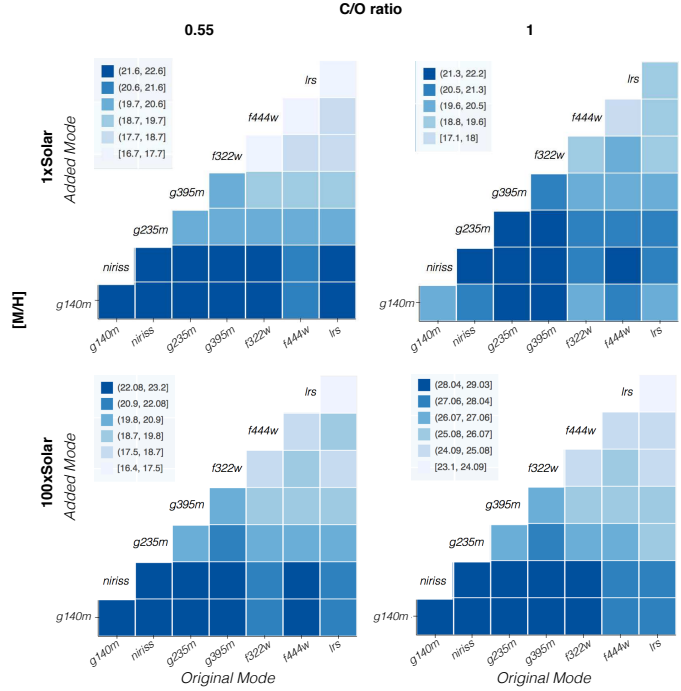}
\caption{Information content maps (in bits) for the T=1800 K planet cases as shown in Figure \ref{fig1} with the addition of a grey cloud at 1 millibar. Rows are different [M/H]'s and columns are different C/O's. Information content is measured in bits and the bins used for each color map is shown next to each panel. Note that each panel has a different color scale. Diagonal elements are two transits in one mode. Observation modes which will maximize observers chances of obtaining the true atmospheric state appear as dark squares in all four panels. \label{fig7}}
\end{figure}

\begin{figure}
\centering
 \includegraphics[angle=0,width=5in]{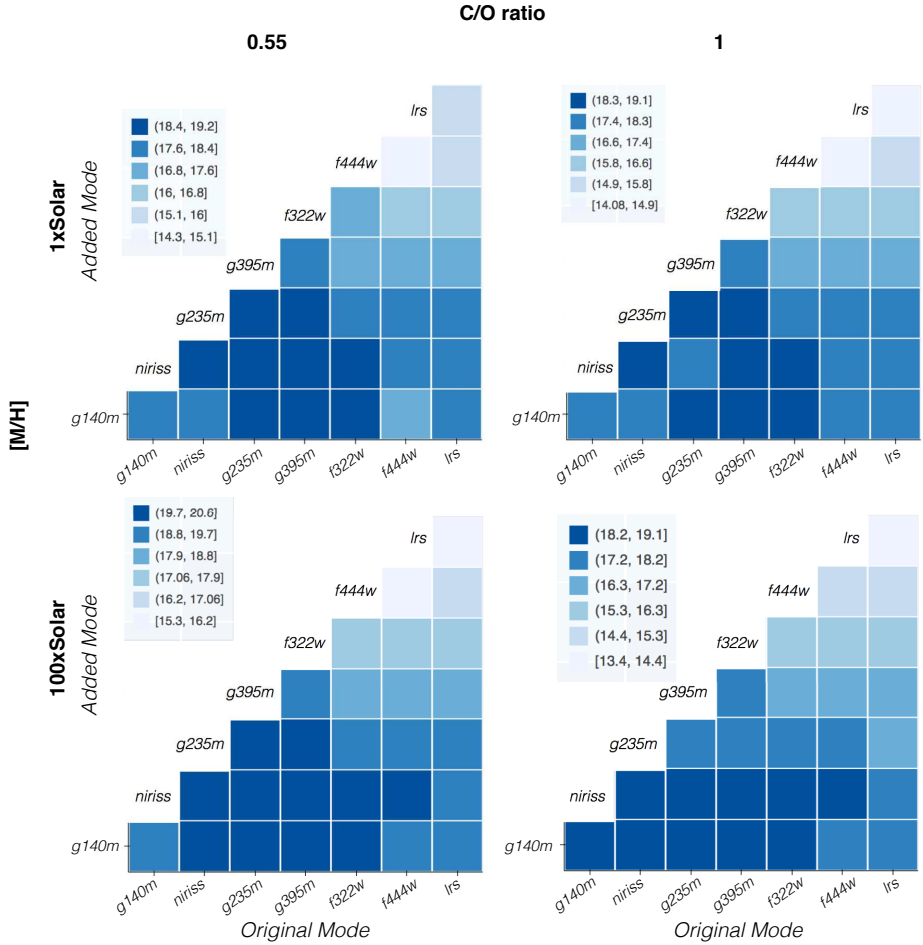}
\caption{Information content maps (in bits) for the same planet as shown in Figure \ref{fig1} but with T=600 K and a grey cloud at 1 millibar. Rows are different [M/H]'s and columns are different C/O's. Information content is measured in bits and the bins used for each color map is shown next to each panel. Note that each panel has a different color scale. Diagonal elements are two transits in one mode. Observation modes which will maximize observers chances of obtaining the true atmospheric state appear as dark squares in all four panels. \label{fig8}}
\end{figure}

The IC maps shown in Figs. \ref{fig7} \& \ref{fig8} can be easily interpreted by finding the combination of modes which, regardless of C/O or [M/H], give the highest information content (i.e. common dark regions in all four panels). We only show the maps for the cases where a 1 mbar cloud has been added because for most combination of modes, the cloudy and cloud-free cases are similar. 

For hotter targets (T=1800 K, Figure~\ref{fig7}), the combination of NIRISS and NIRSpec G395M/H yields the highest information content regardless of temperature, C/O and [M/H]. Other combinations of modes such as 2 transits with NIRISS SOSS, NIRSpec G140M/H+G395M/H, and NIRSpec G140M/H+G235M/H also yield high IC observations. 

The combinations with the lowest IC are 2 transits with MIRI LRS, NIRCam F444, or NIRCam F322. The difference between the highest IC combination and the lowest IC combination is $\sim$4 bits. From Figure~\ref{fig6}, we can estimate this to be about a factor in $\sim$3 degradation in the error on T, C/O, and [M/H]. Meaning, an observation with NIRISS and NIRSpec G395M/H will help constrain these parameters $\sim$3 times better than an observation of two transits with MIRI LRS. 

For cooler targets (T=600 K, Figure~\ref{fig8}), the results are similar. NIRISS and NIRSpec G395M/H have the highest IC while two transits with MIRI LRS, or NIRCam F444 have the lowest. The difference, in bits, between the highest IC and the lowest IC mode is also $\sim$4 bits. Therefore, as with the T=1800 K target, it is predicted that this difference will cause a factor in $\sim$3 degradation in the error on T, C/O, and [M/H].

%%%%%%%%%%%%%%%%%%%%%%%%%%%%%%%%%%%%%%%%%%%%%%%%%%%%%%%%%%%
\subsection{Multimode Analysis}
Beyond two modes, one could pose the question of whether or not adding a third or fourth mode would help tighten the constraint on the atmospheric state. Three modes (NIRISS, NIRSpec G395M/H, \& MIRI LRS) will get you across the entire JWST spectroscopic range, 1-12$\micron$. However, it might be the case that a spectrum from 1-5$\micron$ will give you the same constraints as a spectrum from 1-12$\micron$, because of a saturation in information. If this is the case, it might be more beneficial to increase precision in one mode. To test this, we sequentially add a mode and compute the expected errors on each of the state vector parameters.

\begin{figure}
\centering
 \includegraphics[angle=0,width=4in]{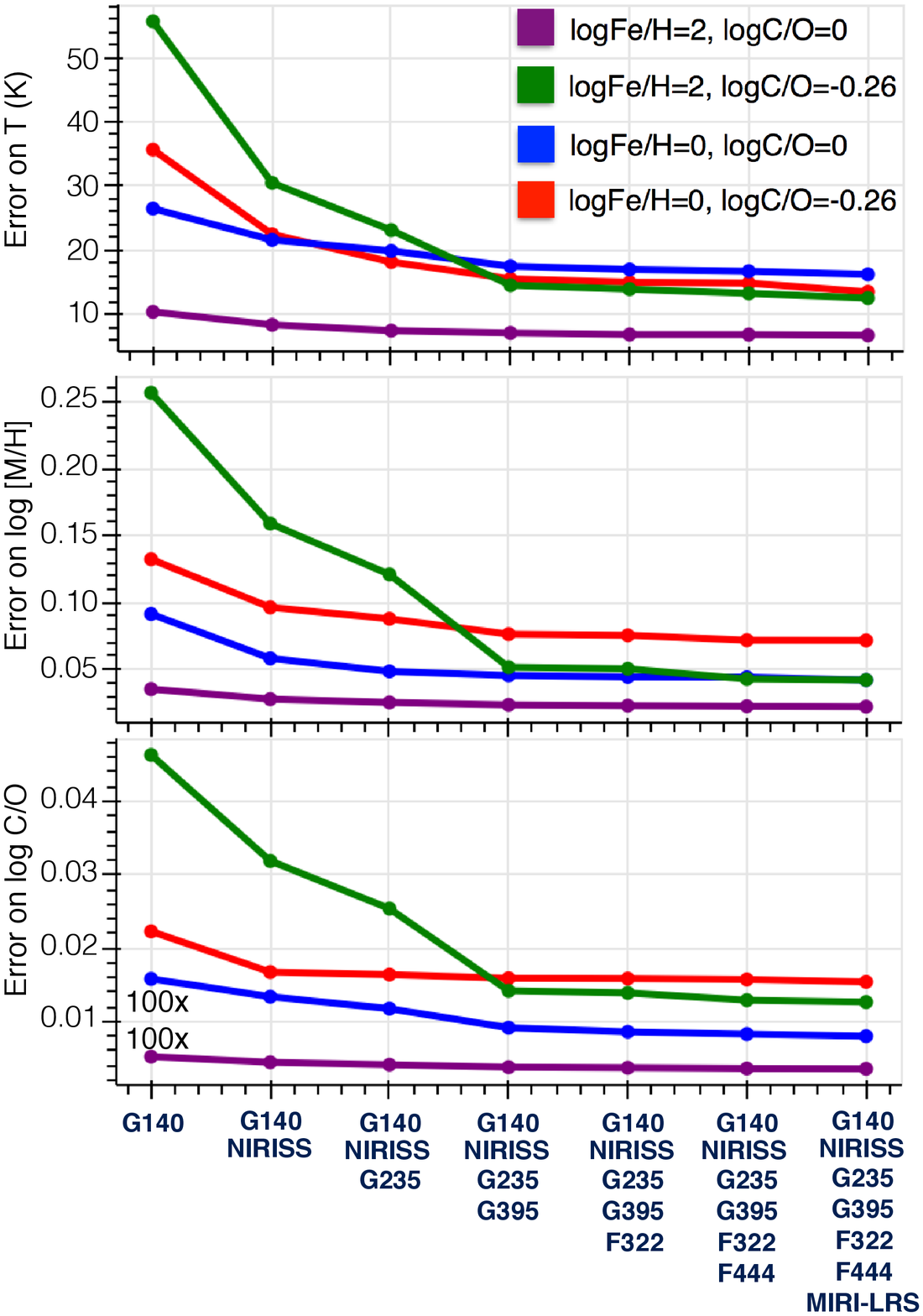}
\caption{Here we show how the errors of the state parameters decrease with the addition of more observation modes. Like the others, this is for the planet system WASP-62b with T$_{eq}$ = 1800 K. The lines represent the four combinations of C/O = 1 or 0.55 and [M/H]=1 or 100$\times$Solar. No clouds have been added to these calculations. \label{fig9}}
\end{figure}

\begin{figure}
\centering
 \includegraphics[angle=0,width=4in]{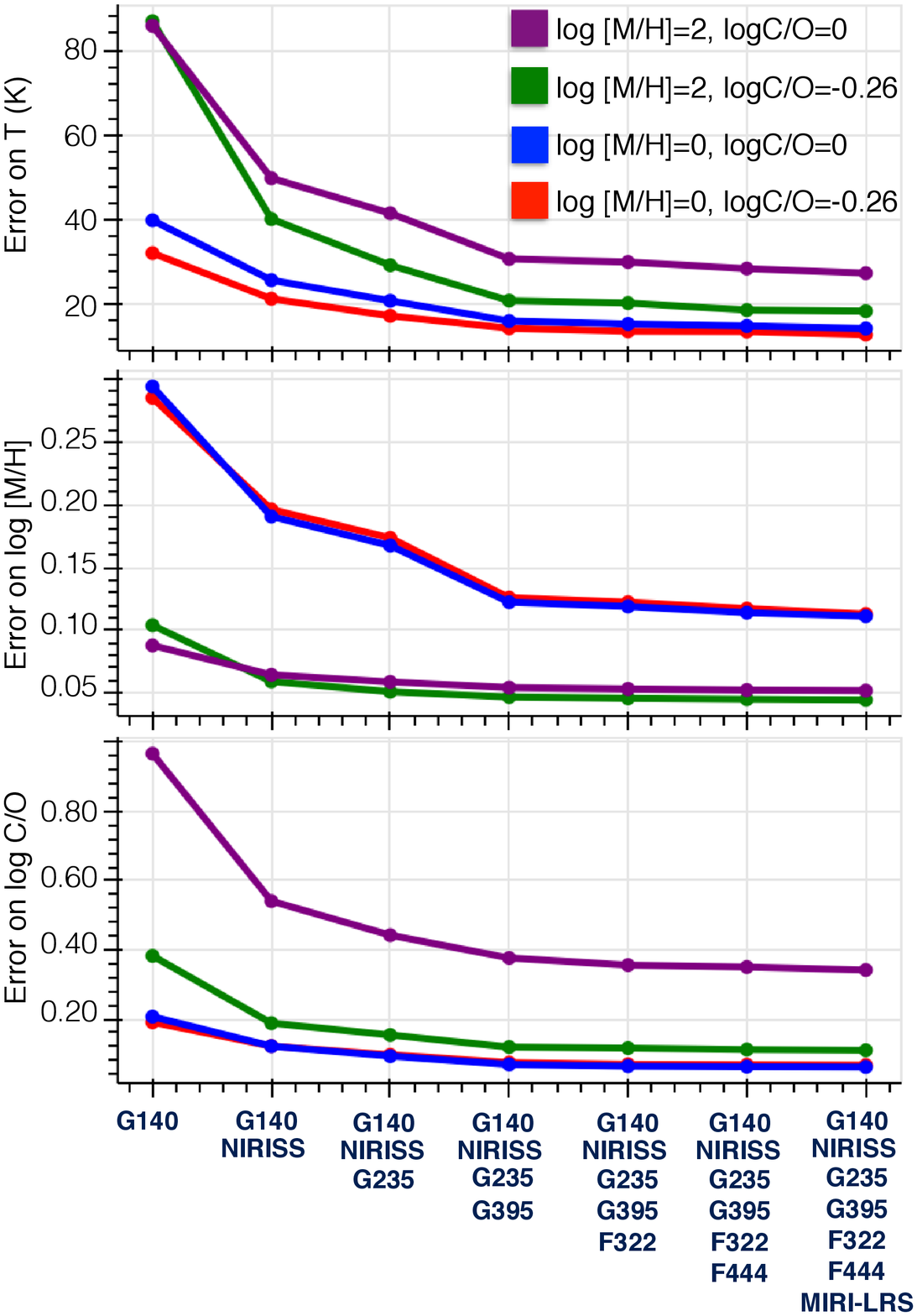}
\caption{Here we show how the errors of the state parameters tightens with the addition of more observation modes. Like the others, this is for the planet system WASP-62b with T$_{eq}$ = 600 K. The lines represent the four combinations of C/O = 1 or 0.55 and [M/H]=1 or 100$\times$Solar. No clouds have been added to these calculations. \label{fig10}}
\end{figure}

Figs. \ref{fig9} \& \ref{fig10} show these results for all chemical configurations and T=1800 K and 600 K, respectively. Starting out, an observation with just NIRSpec G140M/H has relatively poor constraints (for reference, this is equivalent in wavelength space to an observation with the Hubble Space Telescope's WFC3). With the addition of NIRISS SOSS, which effectively improves the precision of the 1-1.7$\micron$ region and adds the 1.7-2.7$\micron$ region, the constraints on T, [M/H] and C/O improve by nearly a factor of 2.  

Adding an observation with NIRSpec G235M/H does not significantly add any new wavelength space, but effectively increases the precision of the observation in the 1.7-3$\micron$ region. This results in only a small improvement of the constraints on the atmospheric state. 

Next, adding in NIRSpec G395M/H effectively expands the wavelength coverage of the observation out to 5$\micron$ and again, in some cases, nearly tightens the constraints of the atmospheric state by a factor of 2. After the addition of NIRSpec G395M/H, adding in NIRCam F322, and F444 does not help because even though the precision of the observation increases in the 3-5$\micron$ region, no more wavelength coverage is being added.

Lastly, the addition of an observation with MIRI LRS does not result in an improvement in the constraint of the atmospheric state. This is unintuitive because the wavelength region is increased out to 12$\micron$. In this case, we have already maximized the total retrievable information from the combined observation of NIRISS SOSS and NIRSpec G395M/H. These results are similar to those found in \citet{gre16}: Adding MIRI LRS did not significantly improve the molecular abundance constraints for transit {\it transmission} spectra. However, observing targets in emission with LRS could significantly improve observations of targets with cloudy transmission spectra \citep{mor15}. We further discuss this in \S5.3. 

%%%%%%%%%%%%%%%%%%%%%%%%%%%%%%%%%%%%%%%%%%%%%%%%%%%%%%%%%%%%%%%%%%
\section{Discussion}
\subsection{Why NIRISS SOSS + NIRSpec G395M/H?}
In transmission, regardless of temperature, C/O, [M/H] or presence of a cloud, a combination of NIRISS+NIRSpec G395M/H yields the highest information content. Intuitively this makes sense because the combination of the two give relatively broad (1-5 $\micron$) wavelength coverage, more so than other combinations.  These wavelength regions are the locations of the spectrum with the highest rate of change with respect to the state vector parameters (the elements of the Jacobian are the highest, Fig \ref{fig2}). This is because these wavelengths cover the prominent absorption features of the metallicity and C/O dependent species (e.g., measuring both oxygen and carbon-bearing compounds). The key 4.5 $\micron$ region probes CO and CO$_2$ which are powerful metallicity indicators (e.g., \citet{mos13}).

NIRCam F322W2 and F444W do not yield high information content results because F444W covers too small of a wavelength region to be useful on its own and F322W cannot capture the presence or absence of carbon based species (CO$_2$ or CO). If the observer knew \emph{a priori} that a carbon species would be present, it could help constrain [M/H] and C/O to get high spectral resolving power observations in that region--especially since NIRSpec G395M's photon-to-electron conversion efficiency starts to drop off after 4$\micron$. Noise simulations will have to be done on a case by case basis to assess this trade off. 

Because the highest information contents per unit observing time are a result of the broad wavelength coverage and not necessarily the precision, the observer could swap out the combination of NIRISS+G395M/H for the NIRSpec prism. The NIRSpec prism was not explored in this paper because it has a very high saturation limit (J$<10.5$). Therefore, some of the best JWST targets will saturate the NIRSpec prism. We conclude that if the target is dim enough, it should be observed with the prism instead of NIRISS SOSS and NIRSpec G395M/H. 

\subsection{Why wavelength coverage over precision?}
We can revisit Eqn. 7 to gain analytical insights into the question of increased precision versus wavelength coverage. The information content is related to the error covariance matrix by, H $\propto \ln|\mathbf{S_e}^{-1}|$. If we were to both expand wavelength coverage, by increasing the number of observation points to $N$, and reduce the error by some factor, $f$, the new information content would go as, H$_{new}$ $\propto N\ln\frac{1}{f^2}$. A second transit in the same mode would only increase the precision by \emph{at most} $\sqrt{2}$, which increase the IC by $\ln2$. This would be less in the case of significant systematic noise. Doubling the number of points, on the other hand, should have a much larger impact on the total IC. This is shown in Figs. \ref{fig8} and \ref{fig9}. An increase in the precision does decrease the overall errors on the state vector parameters, but not as much as an increase in the total number of wavelength points. 

\subsection{Saturation of Information}
The argument of improving parameter constraints by increasing wavelength coverage seems to break down for MIRI LRS, shown in Figs. \ref{fig8} and \ref{fig9}. In these cases, we are expanding wavelength coverage but are seeing no improvement in the errors on the state vector parameters. This is because the 1-5 $\micron$ coverage is already enough to cover the prominent metallicity and C/O sensitive absorbers.  If this were always true, MIRI LRS would never be useful, if combined with NIRISS and NIRSpec G395M. However, there are several important key modeling aspects not included in this investigation.

Firstly, we have not accounted for disequilibrium chemistry (vertical mixing and photochemistry). This will affect planets below 1200K. As an example, the strong 9-12 $\micron$ ammonia feature is very sensitive to the strength of vertical mixing, as demonstrated by brown dwarf observations (e.g., \citet{sau06}). Similarly, obtaining a second methane band at 7.6 $\micron$ will also provide further leverage on the strength of vertical mixing for planets with a carbon reservoir transitioning between CO and methane. Furthermore, methane derived photochemical products like C$_2$H$_2$, C$_2$H$_6$, and C$_2$H$_4$ all have multiple bands long-wards of 5 $\micron$. These longer wavelengths may be diagnostic of photochemistry (e.g., \citet{mos11}).

Second, high temperature condensate clouds might exist in super-hot Jupiter atmospheres \citep{wak16}. The vibrational-mode absorption features of these condensate clouds are only detectable in the 5-14$\micron$ region \citep{wak15, mol16}. Shorter wavelengths are only sensitive to the Mie/Rayleigh scattering slopes/magnitudes of these condensates. The composition at these shorter wavelengths is in turn degenerate with the particle sizes.  Therefore, if this is true, MIRI LRS will be crucial for understanding cloud properties by uniquely identifying their spectral signatures.  

Lastly, emission spectroscopy of will rely heavily on MIRI LRS, especially the coolest planets (e.g, \citet{gre16}). The IC analysis for emission spectroscopy is currently being explored as future work. 

\section{Summary \& Conclusion}
Using a transmission spectra model, we first computed how sensitive the transmission spectrum of different planetary atmospheres are to small changes in temperature, C/O, [M/H] and $\times$R$_p$. These sensitivities make up the Jacobian, which allowed us to compute the information content of each spectrum. The information content, measured in bits, describes how the state of knowledge (relative to the prior) has increased by making a measurement. Therefore, we then sought to find observing schemes that were agnostic to unknown \emph{a priori} knowledge of the atmospheric state. 

First, using only single mode combinations, we computed the IC content as a function of spectral precision on a spectrum for all the JWST exoplanet spectroscopy modes within NIRISS, NIRSpec, NIRCam, and MIRI. We also estimated the errors expected on each of the atmospheric state vector parameters by computing the posterior covariance matrix. We validated this approach, which makes the assumption of Gaussianity of the posterior, against full Monte Carlo based retrievals.  We then extended our IC analysis to combinations of two modes. Finally, we explored the effect of adding more than two modes. Our major conclusions are summarized below are true for transmission spectra only:
\begin{enumerate}
    \item A single observation with NIRISS always yields the highest IC content spectra with the tightest constraints, regardless of temperature, C/O, [M/H], cloud effects or precision. 
    \item A single observation with NIRSpec G140M/H is most susceptible to a loss of information via cloud coverage and should generally not be used alone. 
    \item Generally speaking, expanding wavelength coverage will result in more tightly constrained parameters than observing in the same mode twice. 
    \item An observation with both NIRISS and NIRSpec G395M/H always yields the highest IC content spectra with the tightest constraints, regardless of temperature, C/O, [M/H], cloud effects or precision. NIRSpec prism, which covers the same wavelength space, could also be used in lieu of NIRISS and NIRSpec G395M/H, if the target is dim enough, J$>$10.5. 
    \item Observations that use more than two modes, should be carefully analyzed because sometimes the addition of a third mode results in no gain of information. In these cases, higher precision (more transits) in the original two modes would be more favorable.
\end{enumerate}
JWST is scheduled to launch in October, 2018. After commissioning we will have a much better idea of what the inherent systematics are for each of the exoplanet spectroscopy modes. However, proposals for early release science are due in August, 2017. Therefore, until our knowledge of instruments improve, we can use powerful IC analyses to design optimized observing strategies for future proposals. 

\acknowledgements 
We especially thank Avi Mandell for providing invaluable expertise and insight on JWST noise simulations and useful discussions. We also thank Thomas Greene \& Hannah Wakeford for their helpful comments on the paper. This material is based upon work supported by the National Science Foundation under Grant No. DGE1255832, the Kavli Summer Program in Astrophysics, and NASA Astrobiology Program Early Career Collaboration Award to N.E.B. Any opinions, findings, and conclusions or recommendations expressed in this material are those of the author(s) and do not necessarily reflect the views of the National Science Foundation. M.R.L. acknowledges support provided by NASA through Hubble Fellowship grant 51362 awarded by the Space Telescope Science Institute, which is operated by the Association of Universities for Research in Astronomy, Inc., for NASA, under the contract NAS 5-26555.

%%%%%%%%%%%%%%%%%%%%%%%%%%%%%%%%%%%%%%%%%%%%%%%%%%%%%%%%%%%%%%%%%%
\clearpage

\begin{table}[ht]
\centering
\caption{Instrument modes}
\begin{tabular}{llll}
\hline
\hline
Instrument & Filter & Wavelength Range & Spectral Resolving Power \\
 & & ($\mu$m) & \\
\hline
NIRISS SOSS    &      & 0.6-2.8             &  400-1400  \\
NIRSpec G140M  & F070LP     &  0.9-1.8      &   600-1400    \\
NIRSpec G235M  & F170LP     &  1.70-3.0    &   600-1400      \\
NIRSpec G395M  & F290LP     &  2.9-5  &   600-1400    \\
NIRCam Grism    & F322W2  &  2.5-4.0    &   1000-1770            \\
NIRCam Grism    & F444W   &  3.9-5.0     &   1770-2200            \\
MIRI LRS    &    &  5.0-14             &   100            \\
\hline
\label{tab1}
\end{tabular}
\end{table}
\clearpage
%%%%%%%%%%%%%%%%%%%%%%%%%%%%%%%%%%%%%%%%%%%%%%%%%%%%%%%%%%%%%%%%%%
\clearpage
%%%%%%%%%%%%%%%%%%%%%%%%%%%%%%%%%%%%%%%%%%%%%%%%%%%%%%%%%%%%%%%%%%
\clearpage
%%%%%%%%%%%%%%%%%%%%%%%%%%%%%%%%%%%%%%%%%%%%%%%%%%%%%%%%%%%%%%%%%%
\clearpage
%%%%%%%%%%%%%%%%%%%%%%%%%%%%%%%%%%%%%%%%%%%%%%%%%%%%%%%%%%%%%%%%%%
\clearpage
\clearpage
%%%%%%%%%%%%%%%%%%%%%%%%%%%%%%%%%%%%%%%%%%%%%%%%%%%%%%%%%%%%%%%%%%
\clearpage
%%%%%%%%%%%%%%%%%%%%%%%%%%%%%%%%%%%%%%%%%%%%%%%%%%%%%%%%%%%%%%%%%%
\clearpage
%%%%%%%%%%%%%%%%%%%%%%%%%%%%%%%%%%%%%%%%%%%%%%%%%%%%%%%%%%%%%%%%%%
\clearpage
%%%%%%%%%%%%%%%%%%%%%%%%%%%%%%%%%%%%%%%%%%%%%%%%%%%%%%%%%%%%%%%%%%
\clearpage
%%%%%%%%%%%%%%%%%%%%%%%%%%%%%%%%%%%%%%%%%%%%%%%%%%%%%%%%%%%%%%%%%%

\end{document}